\documentclass[RNAAS]{aastex62}

\newcommand{\rsol}{R_{\odot}}
\newcommand{\lsol}{L_{\odot}}
\newcommand{\msol}{M_{\odot}}
\newcommand{\mearth}{M_{\oplus}}
\newcommand{\rearth}{R_{\oplus}}
\newcommand{\mjup}{M_{\rm Jup}}
\newcommand{\rjup}{R_{\rm Jup}}

\begin{document}

\title{Climate modelling of hypothetical moon-moons in the Kepler-1625b system}

\correspondingauthor{Duncan Forgan}
\email{dhf3@st-andrews.ac.uk}

\author[0000-0003-1175-4388]{Duncan Forgan}
\affiliation{Centre for Exoplanet Science, SUPA, University of St Andrews, North Haugh, St Andrews, KY16 9SS, UK}



\section{Motivation} 

If the exomoon candidate orbiting Kepler-1625b truly exists, it is much more massive than the moons observed in the Solar system \citep{Teachey2017,Teachey2018}.  This exomoon would be sufficiently large to stably host its own satellite.  This has sparked discussion of a new category of celestial object - a moon-moon \citep{Forgan2018c} or submoon \citep{Kollmeier2018}.  

In this Note, I describe initial results of climate modelling of a hypothetical moon-moon in the Kepler-1625b system, calculated using the OBERON code, which jointly computes 1D latitudinal energy balance models for individual worlds alongside the dynamical evolution of the system they inhabit (\citealt{Forgan2016f}, DOI:10.5281/ZENODO.61236).   Both the code and the parameter files used in these runs are available at \url{github.com/dh4gan/oberon}.

\section{System setup} 

\noindent We set up the Kepler-1625b system according to the parameters derived by \citet{Teachey2018}.  The host star, Kepler-1625, has a mass of 1.04 $\msol$ and radius $1\rsol$.  Kepler-1625b is a $10 \mjup$ planet, with radius $1 \rjup$, and orbits the star at a semimajor axis of 0.98 AU, with zero eccentricity.  The exomoon candidate Kepler-1625b-i is 1 Neptune radius (3.8 $\rearth$) in size, and 1 Neptune mass ($17 \mearth$).  It orbits Kepler-1625b with a semimajor axis of 0.02149 AU, and zero eccentricity.

The moon-moon has a mass of 1 $\mearth$, and radius 1$\rearth$, and orbits the exomoon at a distance of 0.000496 AU, beyond the Roche limit and within the dynamical stability limit defined by the exomoon's Hill radius \citep{Domingos2006,Kollmeier2018}. We assume its atmosphere is identical to Earth's. Its initial eccentricity is set to zero, but we find that it takes on a non-zero value of approximately 0.05 during the system's evolution.

We neglect tidal heating and planetary illumination effects, as well as the carbonate-silicate cycle for these preliminary calculations.  Full details of the algorithms used in this calculation are given in \citet{Forgan2016c}.  The simulation is run twice: in the first run, we assume Kepler-1625 is on the main sequence and use main sequence relations to determine its luminosity.  In the second, we use a post-main sequence luminosity of 2.5 $\lsol$.

\section{Results}

\begin{figure}[h!]
\begin{center}
\includegraphics[scale=0.7]{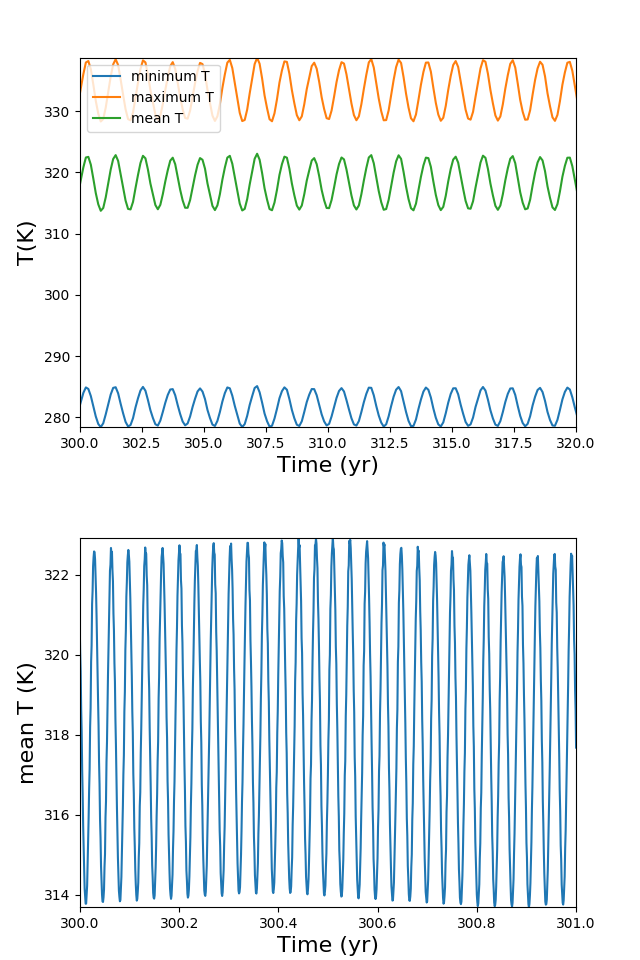}
\caption{Top: The minimum, maximum and mean surface temperature of the moon-moon over a 20 year time interval.  Note that this curve has been downsampled to show temperature fluctuations over the planetary period.  Bottom: The mean temperature over a 1 year time interval.  This has not been downsampled, and reflects the variations on the exomoon candidate's period of 0.03 years. \label{fig:1}}
\end{center}
\end{figure}

\noindent The top panel of Figure \ref{fig:1} shows the maximum, minimum and mean temperature on the surface of the moon-moon after 300 years, while Kepler-1625 remains on the main sequence.  As can be seen, the moon-moon maintains surface temperatures in the range $T= [280 , 340]$ K.  As expected from previous studies, the moon-moon's orbit is also stable.  

Periodogram analysis reveals the temperature has principal amplitudes corresponding to the planetary period about the star of 0.95 years, and the lunar period of 0.03 years.  The climate data is sampled every $10^{-3}$ years, and can be seen in the bottom panel of Figure 1.  The top panel of Figure 1 is downsampled by a factor of 100 to elucidate temperature variations on the planetary period.

Simulations using the post-main sequence luminosity yield an uninhabitable moon-moon, with maximum temperatures exceeding 500 K.

\newpage
\section{Conclusions}

This suggests that moon-moons can indeed be habitable, at least on the timescales simulated here, and while Kepler-1625 was on the main sequence.  Longer simulations are required to ensure that the moon-moon can remain orbitally stable on secular timescales.  We should also note that the non-zero eccentricities exhibited here will produce tidal heating, which may affect its habitability.

Nonetheless, this Note demonstrates that if large exomoons are found in the Universe, this raises not only the prospect of moon-moons, but habitable moon-moons.

\acknowledgments

\bibliographystyle{mnras}
\bibliography{moonmoon}

\end{document}